\documentclass[aps,prl,superscriptaddress,twocolumn]{revtex4} 
\usepackage{graphicx}

\begin{document}
\title{$B_s \to K^{(*)0} \bar K^{(*)0}$ DECAYS:\\ THE GOLDEN CHANNELS FOR NEW
  PHYSICS SEARCHES}

\author{M.~Ciuchini}
\affiliation{Dipartimento di Fisica, Universit\`a di Roma Tre 
and INFN, Sezione di Roma Tre, Via della Vasca Navale 84, I-00146
Roma, Italy}
\author{M.~Pierini}
\affiliation{Department of Physics, University of Wisconsin, Madison,
  WI 53706, USA}
\author{L.~Silvestrini}
\affiliation{Dipartimento di Fisica, Universit\`a di Roma ``La
  Sapienza''  and INFN, 
  Sezione di Roma, P.le A. Moro 2, I-00185 Rome, Italy}

\begin{abstract}
  We point out that time-dependent CP asymmetries in $B_s \to
  K^{*0}\bar K^{*0}$ decays probe the presence of new physics in $b
  \to s$ transitions with an unprecedented theoretical accuracy. We
  show that, contrary to the case of $B_d \to \phi K_S$, it is
  possible to obtain a model-independent prediction for the
  coefficient $S(B_s \to K^{*0}\bar K^{*0})$ in the Standard Model. We
  give an estimate of the experimental precision achievable with the
  next generation of $B$ physics experiments. We also discuss how this
  approach can be extended to the case of $B_s \to \bar K^{*0} K^0$, $B_s
  \to K^{*0} \bar K^0$ and $B_s \to K^0 \bar K^0$ decays and the different
  experimental challenges for these channels.
\end{abstract}

\maketitle

The measurement of CP asymmetries in flavour changing neutral current
processes represents a crucial test of the Standard Model (SM). In
particular, time-dependent CP asymmetries in $b \to s$
penguin-dominated modes are considered among the most sensitive probes
of New Physics (NP)~\cite{pinguinibs}. Measuring these asymmetries is one of the
highlights of the B-factory physics
program~\cite{altreref,babarphiks,bellephiks}. In this context, the study of
$B_d
\to \phi K_S$ has been considered for a long time the {\em golden
  mode} for NP searches in nonleptonic $B$ decays, since it is a pure
penguin~\cite{phikssm}.  Indeed, writing the amplitude in terms of
renormalization group invariant (RGI) parameters, defined in ref.~\cite{burassilv}, one obtains:
\begin{equation}
A(B_d \to \phi K^0) = V^*_{tb}V_{ts}\,P - V^*_{ub}V_{us}\,P^\mathrm{GIM}\,,
\label{eq:phiKamp}
\end{equation}
where $P$ contains penguin contractions of charmed current-current
operators together with the matrix elements of $b \to s$ penguin
operators, while $P^\mathrm{GIM}$ represents the GIM-suppressed
difference of penguin contractions of current-current operators
containing charm and up quarks respectively.

Neglecting the contribution of $P^\mathrm{GIM}$ on the basis of
plausible dynamical arguments, the $B_d \to \phi K_S$ decay is
mediated by a single amplitude, so that under this assumption no
direct CP violation can be produced and the time-dependent CP
asymmetry probes the phase $2\beta$ of $B_d-\bar B_d$ mixing. In
terms of the coefficients $S$ and $C$ of sine and cosine terms in the
time-dependent CP asymmetry, this means that in the SM one expects
$S(B_d \to \phi K^0) = \sin 2\beta$ and $C(B_d \to \phi K^0) = 0$. The
theoretical error associated to this prediction is related to the
ratio of $P^\mathrm{GIM}/P$.

The average of currently available experimental measurements by
BaBar~\cite{babarphiks} and Belle~\cite{bellephiks} gives $S(B_d \to
\phi K^0) = 0.39 \pm 0.18$ and $C(B_d \to \phi K^0) = 0.01 \pm
0.13$~\cite{hfag}. Even though the experimental errors are still
large, it is interesting to observe that the value of $S(B_d \to \phi
K^0)$ deviates from the world average $\sin 2\beta = 0.675 \pm
0.026$~\cite{hfag}. If confirmed in the future with a smaller error,
this measurement might provide a hint of NP in $B$ decays.  On the
other hand, this interpretation should consider the theoretical
uncertanty introduced by neglecting $P^\mathrm{GIM}$. Even though
several model-dependent approaches, based on factorization
or on flavour symmetries, have been proposed in
literature~\cite{deltaSphiksfact,deltaSphikssu3}, a model-independent
evaluation of the error is not available yet. Considering that the
next generation of $B$ physics experiments~\cite{nuoviexp} is expected
to reduce the error on $S(B_d \to \phi K_S)$ down to a few percent,
the lack of a model independent evaluation of the theoretical error is
a strong limitation for a complete and meaningful test of the SM.

In this letter, we propose to overcome this problem by using $B_d \to
K^{(*)0} \bar K^{(*)0}$ decays to predict $S(B_s \to K^{*0} \bar
K^{*0})$ within the SM, including the theoretical error associated to
hadronic uncertanties, in particular to the GIM-suppressed penguin
contractions.  Considering the experimental precision that is expected
at LHCb~\cite{lhcb} and at a future super B-factory~\cite{superb}, we
give a theoretical estimate of the deviation from zero of $S(B_s \to
K^{*0} \bar K^{*0})$ within the SM. This is a crucial ingredient to
search for NP effects in this decay mode: a deviation from zero much
larger than the estimated SM error would be a strong signal of NP.

For a given polarization of the final state~\footnote{Our formulae
  and results apply to any given polarization of the final state. For
  simplicity, we omit the corresponding label. We comment on the
  experimental aspects of measuring vector final states and of
  extracting polarized amplitudes after presenting our method for
  a single polarization.}, we can write the decay amplitude of $B_s \to
K^{*0}\bar K^{*0}$ decays as
\begin{equation}
  A(B_s \to K^{*0} \bar K^{*0}) = -V^*_{tb}V_{ts}\, P_s -
  V^*_{ub}V_{us}\, P_s^\mathrm{GIM}\,,  
\label{eq:KKamp}
\end{equation}
in the same notation of Eq.~(\ref{eq:phiKamp}). Comparing
Eq.~(\ref{eq:KKamp}) to Eq.~(\ref{eq:phiKamp}), it is clear that 
the same NP in $b \to s$ penguins enters both this channel and the
{\it golden mode} $B_d \to \phi K_S$.

From an experimental point of view, the $K^{*0}$ mesons can be
reconstructed as $K^{*0} \to K^+ \pi^-$ and $\bar K^{*0} \to K^-
\pi^+$. Since the final state is a CP eigenstate, it is possible to
measure the CP asymmetry parameters $S$ and $C$ from the
time-dependent study of the tagged decay rates. The information on the
flavour of the decaying $B_s$ is provided by the usual tagging
techniques. Since in this case the $B_s$ meson decays only to charged
tracks directly originating from the vertex, the reconstruction and
vertexing of the $B_s$ mesons should be possible at LHCb, allowing to
measure the parameters of the time-dependent CP asymmetry.

With the same approximation of $B_d \to \phi K_S$, {\it i.e.}\/
neglecting the CKM suppressed contribution of $P_{s,d}^\mathrm{GIM}$,
the SM expectation values for the coefficients of the CP asymmetry are
simply given by $S(B_s \to K^{*0} \bar K^{*0}) = 0$ and $C(B_s \to
K^{*0} \bar K^{*0}) = 0$, as
$$
\lambda_{CP}(B_s \to K^{*0} \bar K^{*0})=e^{2i\beta_s} \frac{A(\bar B_s \to
K^{*0} \bar K^{*0})}{A(B_s \to K^{*0} \bar K^{*0})}=1\,.
$$
This is a null test of the SM, but an estimate of the error
induced by neglecting $P_s^\mathrm{GIM}$ is needed.

The advantage of this mode, with respect to the case of $B_d \to \phi
K^0$, is represented by the possibility of calculating the theoretical
error in a model independent way, using the measurement of $BR(B_s \to
K^{*0} \bar K^{*0})$ and $C(B_s \to K^{*0} \bar K^{*0})$,
together with the information on the order of magnitude of
$P_s^\mathrm{GIM}$ provided by the time-dependent study of $B_d \to K^{*0}
\bar K^{*0}$ decays.

The idea follows the calculation of the error on $\sin 2\beta$ in $B_d
\to J/\psi K^0$ presented in ref.~\cite{cpsbeta}. The expression
for the decay amplitude of $B_d \to K^{*0} \bar K^{*0}$ in the same
notation of Eqs.~(\ref{eq:phiKamp}) and (\ref{eq:KKamp}) is given by
\begin{equation}
A(B_d \to K^{*0} \bar K^{*0}) = -V^*_{tb}V_{td}P_d -
V^*_{ub}V_{ud}P_d^\mathrm{GIM}, 
\label{eq:BdKKamp}
\end{equation}
which is equivalent to Eq.~(\ref{eq:KKamp}), except that
in this case the two combinations of CKM matrix elements have the same
order of magnitude. As a consequence, the sensitivity to
$P_d^\mathrm{GIM}$ is maximal in this case. From the measurement of
the $BR$ and the CP parameters $S$ and $C$, fixing the CKM elements to
their SM values obtained by the UT fit~\cite{utfit}, one can determine
$|P_d|$, $|P_d^\mathrm{GIM}|$, and the relative strong phase
$\delta_{d}$. In the SU(3)-symmetric limit,
$P_d^{\mathrm{(GIM)}}=P_s^{\mathrm{(GIM)}}$ and $\delta_d = \delta_s$.
Imposing these relations, as done in ref.~\cite{hep-ph/9903540}, would
introduce a (difficult to estimate) error associated to SU(3)
breaking~\footnote{In ref.~\cite{matias}, SU(3) was used in
  conjunction with QCD factorization to relate CP violation in $B_s$
  and $B_d$ decays to two kaons. This is another interesting example
  of model-dependent predictions for CP violation in $b \to s$
  penguins.}. To be conservative, we instead allow for a SU(3) breaking
up to 100\%, much larger than any known breaking effect.

Starting from these considerations, the estimate of the theoretical
expectation for the deviation of  $S(B_s \to K^{*0}\bar K^{*0})$ from
zero proceeds through three steps, in analogy to
ref.~\cite{cpsbeta}: i) a fit to determine $P_s$ from $BR(B_s \to
K^{*0} \bar K^{*0})$; ii) a fit of $|P_d|$, $|P_d^\mathrm{GIM}|$, and
$\delta_{d}$ from the experimental values of $BR$, $S$, and $C$ of
$B_d \to K^{*0} \bar K^{*0}$. In the fit, only the solution that gives
$|P_d|$ compatible with $|P_s|$ is considered; iii) a fit of the $B_s
\to K^{*0} \bar K^{*0}$ decay amplitude from the experimental values
of $BR$ and $C$, performed forcing the absolute value $|P_s^\mathrm{GIM}|$ in
the range obtained allowing $100\%$ SU(3) breaking effects around the central
value of $|P_d^\mathrm{GIM}|$. To be conservative, no information from $B_d \to
K^{*0} \bar K^{*0}$ is used for constraining $|P_s|$ and $\delta_{s}$.

Let us first discuss the experimental prospects. Based on these, we
then present an example of how our method might work once the relevant
modes will be measured.  An estimate of the level of precision
reachable at LHCb is difficult to give at this stage and goes beyond
the purpose of this work, since details on the reconstruction of the
LHCb detector are needed. In addition, the lack of measurements of
$B_d \to K^{*0}\bar K^{*0}$ makes any prediction harder.
Nevertheless, few educated assumptions might help us to understand the
order of magnitude of the experimental error on $S(B_s \to K^{*0} \bar
K^{*0})$. We assume that i) LHCb will provide a measurement of the $S$
and $C$ parameters with an error of $\sim 0.02$ (comparable to what is
expected for $B_s \to K^+K^-$); ii) a $5\%$ precision on the decay
rate will be obtained at LHCb or at a super B-factory running at the
$\Upsilon(5S)$ resonance~\cite{Baracchini:2007ei}; iii) a similar
precision will be available for $B_d \to K^{*0} \bar K^{*0}$ rates and
CP asymmetries.  Concerning this last point, it is important to stress
that $B_d \to K^{*0} \bar K^{*0}$ decays are CKM suppressed
with respect to $B_s \to K^{*0} \bar K^{*0}$. Nevertheless, with LHCb
integrating more than two years of data and/or a super B-factory
integrating $> 30$ ab$^{-1}$, there should be no limitation given by
the available statistics. For the central values, we assume that they
lie in the ballpark of the calculation of ref.~\cite{benekeVV}, but we
have checked that larger values of the BR's give similar results. The
values we assume are summarized in
Tab.~\ref{tab:assumedvalues}~\footnote{During the review process of
  our manuscript, the BaBar collaboration published a first
  measurement of $BR(B_d \to K^{*0} \bar K^{*0})=(4.9^{+1.6}_{-1.3}\pm
  0.5)\cdot 10^{-7}$, in agreement with the value we assumed, altough
  with a large uncertainty~\cite{arXiv:0708.2248}.}.

\begin{figure}[htb!]
\begin{center}
\includegraphics[width=0.35\textwidth]{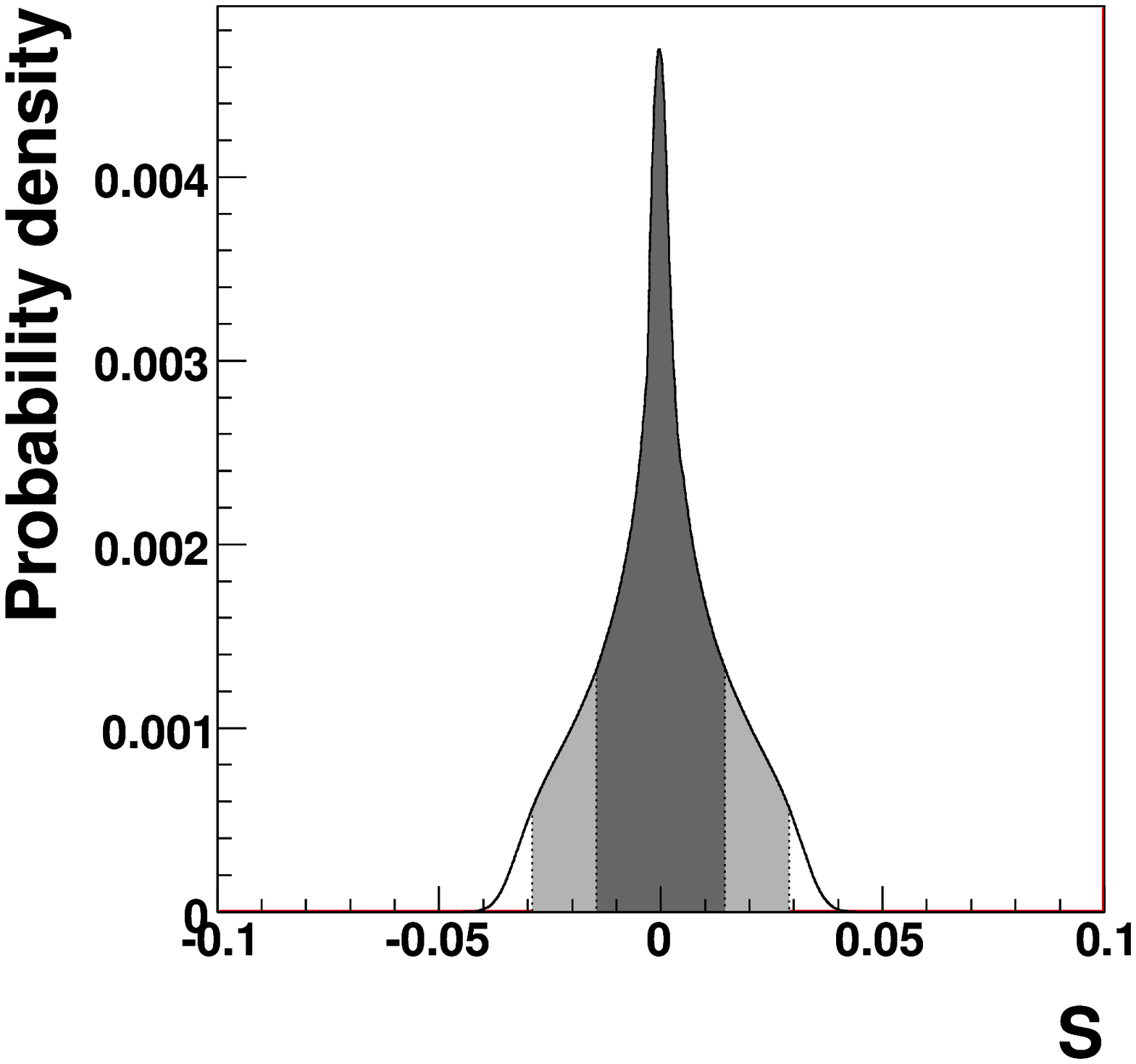}
\includegraphics[width=0.35\textwidth]{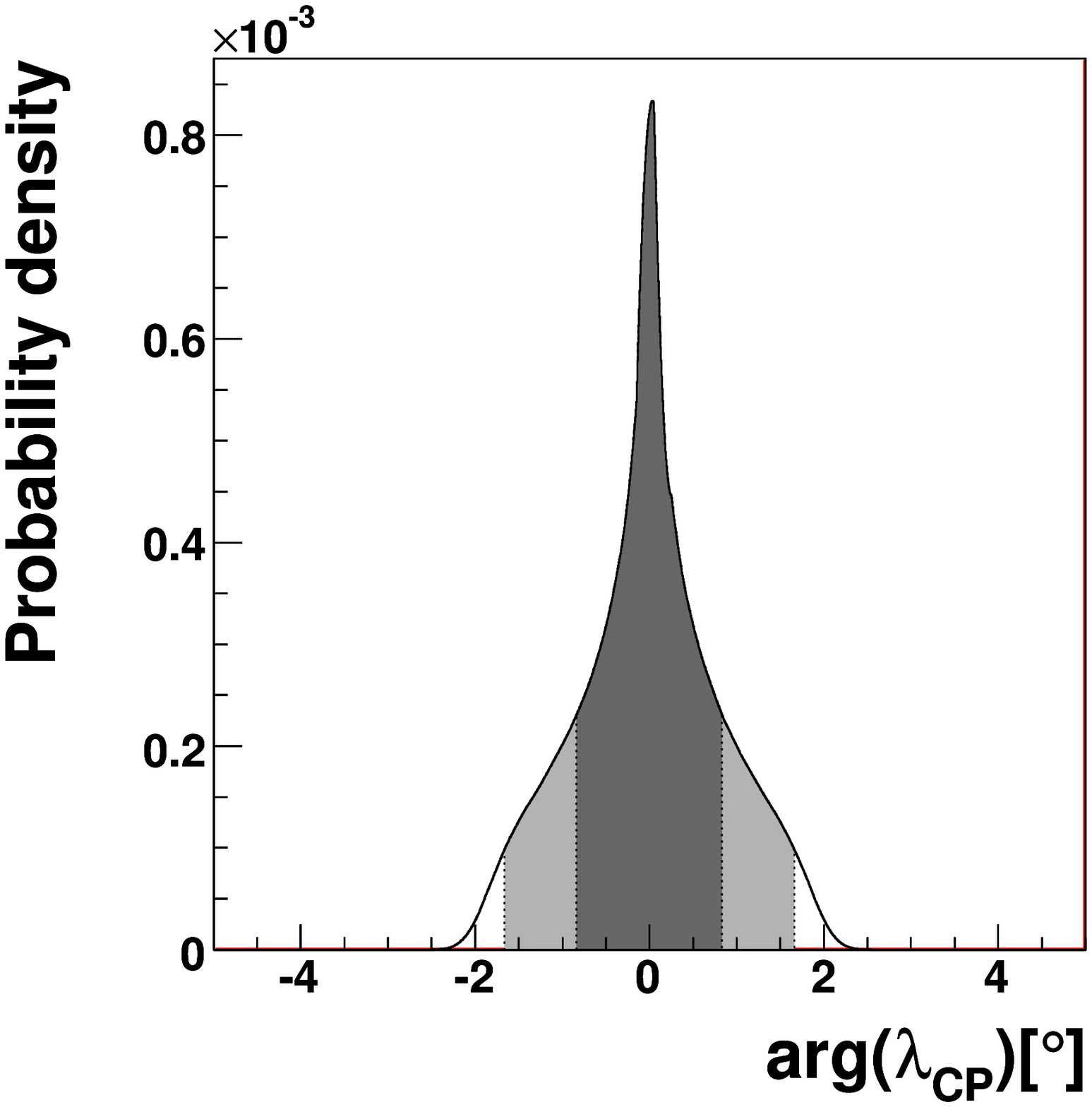}
\caption{%
  Probability density function for $S$ (top)
  and $\arg(\lambda_{CP})$ (bottom) for the decay
  $B_s \to K^{*0}\bar K^{*0}$
  obtained with the procedure detailed in the
  text, using the input values given in
  Tab.~\protect\ref{tab:assumedvalues}.}
\label{fig:d2b}
\end{center}
\end{figure}

We now give an example of the precision we might expect on the
theoretical prediction of $S(B_s \to K^{*0}\bar K^{*0})$ using the
numbers given above.  Using our method we obtain the distribution of
$S(B_s \to K^{*0}\bar K^{*0})$ shown in Fig.~\ref{fig:d2b}, with an
RMS of $0.015$. This corresponds to a theoretical error on
$\arg(\lambda_{CP}(B_s \to K^{*0}\bar K^{*0}))$ of
$0.8^\circ$. Clearly this estimate is only illustrative as it is based
on the values in Tab.~\ref{tab:assumedvalues} inspired by
factorization models. Once data will be available, however, the method
will provide an estimate independent of any theoretical model.

\begin{table}[!tb]
\begin{center}
\begin{tabular}{@{}cccc}
\hline\hline  
channel & BR & S & C \\
\hline
$B_s \to K^{*0}\bar K^{*0}$ &
$(11.8 \pm 0.6)10^{-6}$ & $-0.07 \pm 0.02$ & $0.01
\pm 0.02$\\ 
\hline
$B_d \to K^{*0}\bar K^{*0}$ &
$(5.00 \pm 0.25)10^{-7}$ & $-0.12 \pm 0.02$ & $0.13 \pm
0.02$ \\ 
\hline\hline  
\end{tabular}
\end{center}
\caption{Input values used to estimate the precision on the
determination of $\arg(\lambda_{CP})$.}
\label{tab:assumedvalues}
\end{table}

The presence of multiple polarizations in the $K^* \bar K^*$ final state
does not change the idea we propose here, but it has a practical
impact on the analysis strategy. Our procedure can be followed for each
polarization (longitudinal or transverse), taking into account the relative
minus sign in the CP eigenvalue. Experimentally, using an angular
analysis it is possible to separate the different contributions and
independently determine rates and CP parameters~\footnote{In case
large correlations are observed among the three rates and the three sets
of CP parameters, the theoretical analysis will have to take them
into account.}.

In terms of the experimental fit, the separation of the three
polarizations requires to add the angular distribution of the 
final state particles to the maximum likelihood fit, as it was done
for example in the $B_d\to \rho^+\rho^-$ time-dependent analyses of
BaBar~\cite{arXiv:0705.2157}.

From a practical point of view, the presence of three polarizations 
helps to increase the experimental precision, with respect to the case
of a single polarization.
In fact, the measurement is also sensitive to interference terms
among the different polarizations, as for $B_d\to J/\psi K^*$ time-dependent
analyses performed at the B-factories~\cite{hep-ex/0411016,hep-ex/0504030}.
One can define eleven parameters describing the
complex $B_d$ decay amplitudes for the three polarization states
and their CP conjugates, up to an arbitrary global phase. In terms of
these eleven parameters one can compute three sets of rates, $S$ and $C$
coefficients (one for each polarization). Since the number of unknowns
is smaller than the number of observables, the presence of three
polarizations in the final state will improve the precision of the analysis.
The same procedure can be used for $B_s$ decays.

In principle, the same approach can also be applied to $B_s \to K^{*0}
\bar K^{0}$, $B_s \to \bar K^{*0} K^{0}$ and $B_s \to K^{0} \bar K^{0}$ decays,
with the caveat that the strategy
has to change in order to face the different experimental challenges.

For $B_s \to K^0 \bar K^0$, the measurement of the BR should be
possible at LHCb or at a super B-factory. On the other hand, the
time-dependent CP parameters $S$ and $C$ cannot be measured, since the
extrapolation of the $B_s$ vertex from the flight direction of two
$K_S$ does not seem possible at LHCb, while a B-factory has not enough
vertex resolution to follow the fast oscillations of $B_s$
mesons~\cite{Baracchini:2007ei}.  Nevertheless, it is still possible
to obtain a determination of $\lambda_{CP}$, measuring the tagged
decay rates for $\Delta t>0$ and $\Delta t<0$.  The sign of $\Delta t$
can be measured at a super B-factory, using the $K_S$ flight direction
to determine the $B$ vertex~\cite{k0pi0exp}. Using a full Monte Carlo
simulation, it was shown that it is possible to measure
$\arg(\lambda_{CP}(B_s \to K^0 \bar K^0))$ with an experimental error
less than $20^\circ$~\cite{Baracchini:2007ei}.  The actual error could
be even smaller, if the improvement of the vertexing detector (due to
the use of a layer zero of the silicon detector close to the beam
pipe) will allow to separate primary and secondary vertices on $B \to
D X$ decays~\cite{Bona:2007qt}, strongly reducing the background
contamination~\footnote{Concerning the tagging at
  the $\Upsilon(5S)$, the training of a tagging algorithm for the
  $B_s$ will face new challenges, as both $K$ and $\bar K$ can be
  produced in a $B_s$ decay. One could cope with this problem using, for
  instance, the difference in the momentum of the kaons coming from
  the $b$ quark decay and of the kaons coming from the cascade of
  $D_s$ mesons produced by the $B_s$.  A detailed estimation of the
  change in performances goes beyond the purpose of this paper. We use
  the current $B$-factory performance as a crude estimate of the
  tagging efficiency.}. At the same time, the RMS for
$\arg(\lambda_{CP})$ expected in the SM, taking into account
$P^\mathrm{GIM}_s$ with our method, should be at the level of
$4^\circ$~\cite{Baracchini:2007ei}.

The case of $B_s \to K^{*0} \bar K^{0}$ and $B_s \to \bar K^{*0} K^{0}$
is more similar to $B_s \to K^{*0} \bar K^{*0}$. The
main difference in this case is that there are two
different particles in the final state. As a consequence, the number
of hadronic parameters to determine is twice the number of hadronic
parameters for a single polarization in $B_s \to K^{*0} \bar K^{*0}$ modes.  On
the other hand, the number of experimental observables is larger.
Reconstructing $B_s \to K^{*0} \bar K^0$ and $\bar B_s \to K^{*0} \bar
K^0$ ($B_s \to K^{0}\bar K^{*0}$ and $\bar B_s \to K^{0}\bar K^{*0}$)
from $K^+ \pi^- K_S$ ($K^- \pi^+ K_S$) final states it is possible to
measure CP violating effects~\cite{dunietz}, which provide four
observables ($S$, $C$, $\bar S$, and $\bar C$) in addition to two
decay rates.  It will be possible to use $B_s \to K^{*0} \bar K^0$ and
$B_s \to K^{0}\bar K^{*0}$ to obtain two null tests of the SM,
using the upper values on the two $P^\mathrm{GIM}_s$ contributions
obtained from $B_d \to K^{*0} \bar K^0$ and $B_d \to K^{0}\bar K^{*0}$
respectively.

To summarize, we have proposed a new strategy to look for NP in $b \to
s$ penguins without relying on model-dependent estimates of the
hadronic uncertainties. The new \emph{golden channel} we suggest is
$B_s \to K^{(*)0} \bar K^{(*)0}$. We claim that the SM pollution in
the null tests of the SM from time-dependent CP asymmetries in this
\emph{golden channel} can be controlled with a high accuracy in a
model-independent way. The key
observation is that, even allowing for SU(3) breaking effects of
$\mathcal{O}(1)$, using the experimental information on the
SU(3)-related channel $B_d \to K^{(*)0} \bar K^{(*)0}$ it is possible
to put a strong constraint on the polluting CKM-suppressed penguin
amplitude. The most promising channel seems to be $B_s \to K^{*0} \bar
K^{*0}$, which can be reconstructed from four charged tracks in the final
state and should be easily accessible at LHCb, together with the $B_d$ decay to
the same final state (which can already be studied with the full dataset
collected by BaBar and Belle).
Pseudoscalar-vector and pseudoscalar-pseudoscalar final states imply the
presence of $K_S$ mesons, making the analysis harder in the environment of a
hadron collider. In this respect, a super B-factory would play a very
important role.
\\

We warmly thank Y.~Xie for pointing out to us an embarassing error in
the previous version of this paper. Fortunately correcting this error
further strengthens our proposal.
\\

We aknowledge partial support from RTN European contracts
MRTN-CT-2004-503369 ``The Quest for Unification'', MRTN-CT-2006-035482
``FLAVIAnet'' and MRTN-CT-2006-035505 ``Heptools''.

\end{document}